\documentclass[cits]{PoS}
\usepackage{amssymb, graphicx, latexsym, collref, epsfig, amsmath, bbm, rotating}
\newcommand{\Z}{\mathbb{Z}}

\newcommand{\U}{\mathrm{U}}

\newcommand{\SPol}{S_{\mbox{\tiny{Pol}}}}

\newcommand{\Vext}{V_{\mbox{\tiny{ext}}}}
\newcommand{\QNG}{Q_{\mbox{\tiny{NG}}}}
\newcommand{\Rmin}{R_{\mbox{\tiny{min}}}}

\newcommand{\re}{{\rm{Re}}}

\newcommand{\redchisq}{\chi^2_{\tiny\mbox{red}}}
\newcommand{\eq}{\begin{equation}}
\newcommand{\en}{\end{equation}}
\newcommand{\eqar}{\begin{eqnarray}}
\newcommand{\enar}{\end{eqnarray}}

\title{Effective string description of the interquark potential in the 3D $\U(1)$
lattice gauge theory}

\ShortTitle{Effective string in the 3D $\U(1)$ gauge theory}

\author{\speaker{Davide Vadacchino}, Michele Caselle\\
        Dipartimento di Fisica Teorica, Universit\`a di Torino\\
        and Istituto Nazionale di Fisica Nucleare, sezione di Torino,\\
        Via Pietro Giuria 1, I-10125 Torino, Italy\\
        E-mail: \email{vadacchi@to.infn.it}, \email{caselle@to.infn.it}}

\author{Marco Panero\\
       Instituto de F\'{\i}sica T\'eorica UAM/CSIC, Universidad Aut\'onoma de Madrid\\
       Calle Nicol\'as Cabrera 13-15, Cantoblanco E-28049 Madrid, Spain\\
       E-mail: \email{marco.panero@inv.uam.es}}
   
\author{Roberto Pellegrini\\
       Physics Department, Swansea University,\\
       Singleton Park, Swansea SA2 8PP, UK\\
       E-mail: \email{ropelleg@to.infn.it}}

\abstract{The $\U(1)$ lattice gauge theory in three dimensions is a perfect laboratory to study the properties of the confining string. On the one hand, thanks to the mapping to a Coulomb gas of monopoles, the confining properties of the model can be studied semi-classically. On the other hand, high-precision numerical estimates of Polyakov loop correlators can be obtained via a duality map to a spin model. This allowed us to perform high-precision tests of the universal behavior of the effective string and to find macroscopic deviations with respect to the expected Nambu-Goto predictions. These corrections could be fitted with very good precision including a contribution (which is consistent with Lorentz symmetry) proportional to the square of the extrinsic curvature in the effective string action, as originally suggested by Polyakov. Performing our analysis at different values of $\beta$ we were able to show that this term scales as expected by Polyakov's solution and dominates in the continuum. We also discuss the interplay between the extrinsic curvature contribution and the boundary correction induced by the Polyakov loops.
\vspace{1cm}
\begin{flushright}
IFT-UAM/CSIC-14-109
\end{flushright}
}

\FullConference{The 32nd International Symposium on Lattice Field Theory,\\
                23-28 June, 2014\\
                Columbia University New York, NY}

\begin{document}

\section{$\U(1)$ gauge theory in three spacetime dimensions}
\label{sect:U1}

In our recent work reported in ref.~\cite{Caselle:2014eka}, we studied numerically the $\U(1)$ gauge theory in three spacetime dimensions. The Wilson discretization for the action of this theory is
\eq
\label{Wilson_action}
\beta \sum_{x \in \Lambda} \sum_{1 \le \mu < \nu \le 3} \left[1 - \re \, U_{\mu\nu}(x)\right], \;\;\; \mbox{with} \;\;\; \beta=\frac{1}{a e^2},
\en
where $\Lambda$ is a cubic lattice of spacing $a$, $U_{\mu\nu}(x)$ is the plaquette, and $U_\mu(x) = \exp \left[ i a A_\mu \left( x + a \hat{\mu}/2 \right) \right]$.
A semi-classical analysis shows that this theory is confining for all values of $\beta$, and that it reduces to a theory of free massive scalars for $ \beta\gg 1$~\cite{Polyakov:1976fu, Gopfert:1981er}. In this limit the mass of the lightest glueball ($m_0$) and 
the string tension ($\sigma$) are given by
\begin{equation}
m_0 a = c_0\sqrt{8 \pi^2 \beta}e^{-\pi^2\beta v(0)},\;\;\; \sigma a^2 \geq
\frac{c_{\sigma}}{\sqrt{2\pi^2\beta}} e^{-\pi^2\beta v(0)},
\end{equation}
with $c_0=1$ and $c_{\sigma}=8$. In agreement with previous 
numerical studies~\cite{Loan:2002ej}, we find that the string tension saturates this bound and that both $c_0$ and $c_1$ are affected 
by the semi-classical approximation, changing their values in the continuum. 
Nevertheless, both $m_0$ and $\sigma$ are positive for all $\beta$, so the model is confining. At finite lattice spacing, the $m_0/\sqrt{\sigma}$ ratio is given by
\begin{equation}
\label{ratio_m0}
\frac{m_0}{\sqrt{\sigma}} = \frac{2 c_0}{\sqrt{c_\sigma}} 
(2\pi^2\beta)^{3/4} e^{-\pi^2 v(0)\beta/2},
\end{equation}
so it can be set to any value by tuning $\beta$.

An exact duality transformation of this theory maps its partition function to
\eq
\label{dual_partition_function}
Z = \sum_{ \{ {}^\star s \in \Z \} } \prod_{\mbox{\tiny{links}}} I_{|d {}^\star s|} (\beta),
\en
which defines a globally $\mathbb{Z}$-invariant model, with integer-valued ${}^\star s$ variables, residing on dual sites. In eq.~(\ref{dual_partition_function}), $I_\alpha(z)$ is a modified Bessel function of the first kind, the product is taken over the elementary bonds of the dual lattice, and $d {}^\star s$ denotes the difference of ${}^\star s$ variables at the ends of a bond. In four dimensions, this duality yields a model with local $\mathbb{Z}$ symmetry~\cite{Zach:1997yz, Panero:2005iu, Panero:2004zq, Cobanera:2011wn, Mercado:2013ola}. 

This duality maps the original gauge theory to a spin model, making it is faster and easier to simulate. Furthermore, it is easy to add two opposite probe charges at distance $R$: this results into a modified partition function
\begin{equation}
Z_R  = \sum_{ \{ {}^\star s \in \Z \} } \prod_{\mbox{\tiny{links}}} I_{|d {}^\star s+ {}^\star n|} (\beta),
\end{equation}
where ${}^\star n$ is a 1-form with values in $\Z$, which is non-vanishing on a set of 
links dual to a surface bounded by the worldlines of the original charges. Thus, the two-point Polyakov loop correlator becomes
\begin{equation}
\label{Pol_correlator_in_dual}
\langle P^\star (R) P(0) \rangle = \frac{Z_R}{Z} .
\end{equation}
Using the ``snake algorithm''~\cite{deForcrand:2000fi} and a 
hierarchical update scheme~\cite{Caselle:2002ah}, these correlators can be evaluated to high precision, even at large $R$: the duality transformation allows one to bypass the exponential decay of the signal-to-noise ratio. Denoting the length of (the separation between) the Polyakov loops in units of the lattice spacing as $n_t$ ($n_R$), one gets
\begin{equation}
\label{snake}
\langle P^\star (R) P(0) \rangle = \prod_{i=0}^{n_R n_t -1} \frac{Z^{(i+1)}}{Z^{(i)}} ,
\end{equation}
where the partition function in the numerator differs from that in the denominator by just 
one ${}^\star n$ on a bond of the dual lattice: thus the computation of the correlator is 
reduced to the computation of $n_t$ local quantities in independent simulations.

The dual formulation also offers insight into the confinement mechanism at work. As shown in ref.~\cite{Polyakov:1976fu}, confinement is driven by monopole condensation, as advocated in the dual superconductor picture of QCD. For this theory, however, Polyakov~\cite{Polyakov:1996nc} even suggested an \emph{Ansatz} for the effective action which should describe the confining string, and for the dependence of 
its couplings on the lightest glueball mass and on the coupling of the original gauge theory. In the cited work it is pointed out that the action should be
 \eq
\SPol= c_1 e^2 m_0 \int d^2\xi \sqrt{g}~ + c_2 \frac{e^2}{m_0} \int d^2\xi \sqrt{g} K^2~.
 \label{Polyakov}
 \en
Identifying $c_1$ and $c_2$ with $\sigma$ and $\alpha$ one finds $\sqrt{\sigma/\alpha} \sim m_0$. 

This action was first proposed as a model for fluid membranes~\cite{Hel85, Pel85, Forster1986115} and later as a way to stabilize the Nambu-Goto action~\cite{Polyakov:1986cs, Kleinert:1986bk}. Its contribution to the interquark potential was first computed in the limit of a large number of dimensions $D$~\cite{Braaten:1986bz}, and then also for $D$ generic~\cite{German:1989vk}. The corrections it induces onto higher-order terms of the spectrum have been recently calculated~\cite{Ambjorn:2014rwa}. A discussion of the implications of this action for the static interquark potential in the 3D $\U(1)$ gauge theory is presented in a companion contribution~\cite{Caselle_Lattice_2014}.

\section{Numerical evidence for a rigid-string contribution}
\label{sect:results}

We performed a set of simulations on the dually transformed $\U(1)$ lattice model with a local-update Metropolis algorithm. Here and in the following we will mostly use lattice units, setting $a=1$. The computations were done on lattices of size $L^2\times N_t$, with $L=N_t$ ranging from $64$ to $128$, at five values of $\beta$, from $1.7$ to $2.4$, see tab.~\ref{tab:simul}: this setup enabled us to explore a large range of $\sigma$ and $m_0$ values. Note that we always chose $L>10/\sqrt{\sigma}$ and $L>10/m_0$, to suppress finite-volume corrections.

We calculated the ratio of Polyakov loop correlators at distances differing by one lattice spacing, and constructed the quantity
\eq
Q(R) = -\frac{1}{N_t}\ln{\frac{G(R+1)}{G(R)}} = V(R+1)-V(R),
\en
where $G(R)=\langle P^\star(x) P(x+R) \rangle$, $P(x)$ is the Polyakov loop through the 
site $x$ and $V(R)$ denotes the potential between two static sources at distance $R$. 
$Q(R)$ relates the effective string action to the static interquark potential, removing constant and perimeter terms. Using the algorithm described above, we evaluated $Q(R)$ to high precision for several values of $\beta$ and for $1/\sqrt{\sigma}<R<L/2$. 

The data was first fitted with the standard Nambu-Goto effective string expectation\footnote{We performed our fits using the NG expression to all orders 
in the $1/R$ expansion, as in eq.~(\ref{NGfit1}), which is equivalent, within errors, to its truncation at order $O(R^{-3})$.}
\eq
\QNG(R) = \sigma \left[ \sqrt{(R+1)^2 - \frac{\pi}{12 \sigma}} - \sqrt{R^2 - \frac{\pi}{12 \sigma}}\right].
\label{NGfit1}
\en
$Q(R)$ was fitted to the data (using the string tension $\sigma$ as the fitted parameter), for $R$ from $\Rmin$ to $R=L/2$. $\Rmin$ was increased, starting from $1/\sqrt{\sigma}$, until a reduced $\chi^2$ of order $1$ could be obtained, yielding the results reported in the second column of tab.~\ref{tab:simul}. For $\beta<2$ a good fit could be obtained for small $\Rmin$, but for $\beta \geq 2$, the minimal $R$ had to be increased to larger and larger values. The deviations from the Nambu-Goto behavior is manifest in fig.~\ref{fig:fig2}, where the differences $[Q(R)-\QNG(R)]$ (using the asymptotic values of $\sigma$ 
from tab.~\ref{tab:simul} for $\QNG(R)$) are shown.

It is well-known that Lorentz invariance sets strong constraints onto the effective string action~\cite{Aharony:2011gb}. For the 3D $\U(1)$ theory, Polyakov~\cite{Polyakov:1996nc} presented an argument for the existence of a mapping between gauge and string degrees of freedom, and suggested the form of the action describing the string dynamics, including a term
proportional to the square of the extrinsic curvature of the worldsheet. The resulting string is often referred to as the ``rigid string''. Here we follow the analysis performed in ref.~\cite{Caselle_Lattice_2014}, from which we can infer 
that the shape of $Q(r)$ allowed by Lorentz invariance, computed to the leading order in the extrinsic curvature term, is
\eq
Q(R) = \QNG(R) + Q_r(R) + Q_b(R)
\en
with
\eq
 Q_b(R) = - \frac{b_2\pi^3}{60}\left[ \frac{1}{(R+1)^4}-\frac{1}{R^4} \right],\quad
 Q_r(R) =  -\frac{m}{2\pi}\sum_{n=1}^{\infty} \frac{K_1\left(2 n m (R+1)\right)-K_1(2 n m R)}{n},
\en
for $m=\sqrt{\sigma/(2\alpha)}$. $Q_b(R)$ is a boundary contribution and $Q_r(R)$ is the contribution from the extrinsic curvature term in the effective string action. Including the next-to-leading-order (NLO) contribution in the extrinsic curvature term would correspond to replacing $Q_r(R)$ with 
\eq
Q_r'(R) = Q_r(R) + \frac{21}{20 m \sigma}\left(\frac{\pi}{24}\right)^2\left[ \frac{1}{(R+1)^4}-\frac{1}{R^4} \right].
\label{eq:qrprime}
\en

The $[Q(R)-\QNG(R)]$ differences were at first fitted to the boundary correction $Q_b(R)$ alone. However, reasonable values of $\redchisq$ could only 
be reached for very large $\Rmin$, and the resulting values for $b_2$ were inconsistent with the expected scaling behavior. $b_2$ should scale as $\sigma^{-3/2}$, but
we found that $b_2\sigma^{3/2}$ increases from $0.033(3)$ (for $\beta=1.7$) to $0.62(6)$ (for $\beta=2.4$). To rule out the possibility that the deviations could be due solely to a boundary term, we performed a two-parameter fit of the $[Q(R)-\QNG(R)]$ differences 
to a correction term with a free exponent $b$,
\eq
Q_b'(R) = k \left[ \frac{1}{(R+1)^b}-\frac{1}{R^b} \right].
\en
In this case, at every value of the coupling that we investigated, the results of the fits for $b$ ranged between 2 and 3 and reasonable $\redchisq$ values could only be obtained for very large $\Rmin$, resulting in $k$ values essentially compatible with zero, within their uncertainties. A boundary-type correction was therefore rejected as the explanation for the observed deviations.

\begin{figure}[-t]
\centerline{\includegraphics[width=\textwidth]{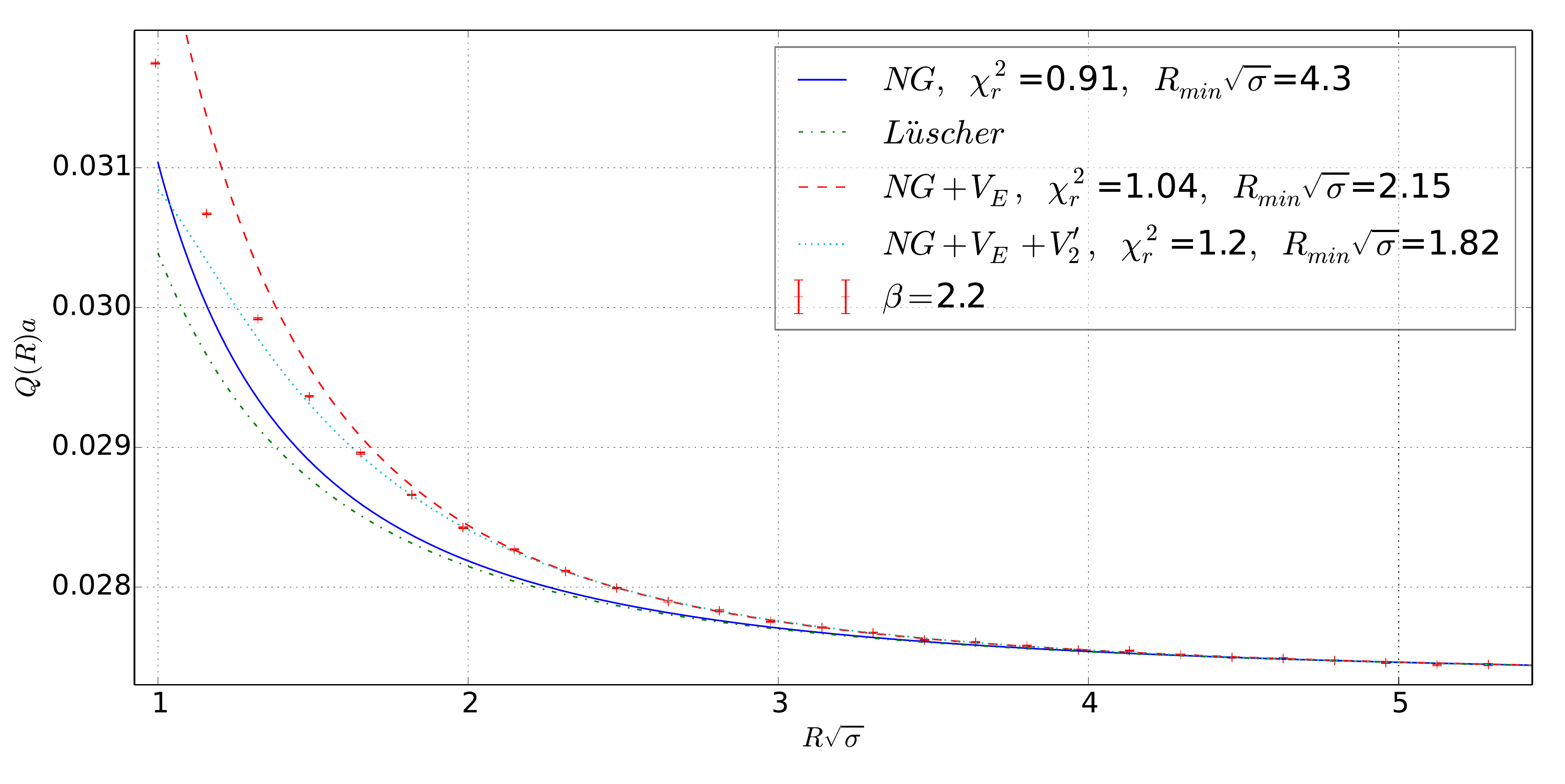}}
\caption{Data and best-fit curves for $\beta=2.2$ with $\QNG$, $\QNG+$ L\"uscher,$\Vext$, and $\Vext+V'_2$. The last two are two-parameter ($\sigma$ and $m$) fits.\label{fig:fig1}}
\end{figure}

\begin{figure}[-t]
\centerline{\includegraphics[width=\textwidth]{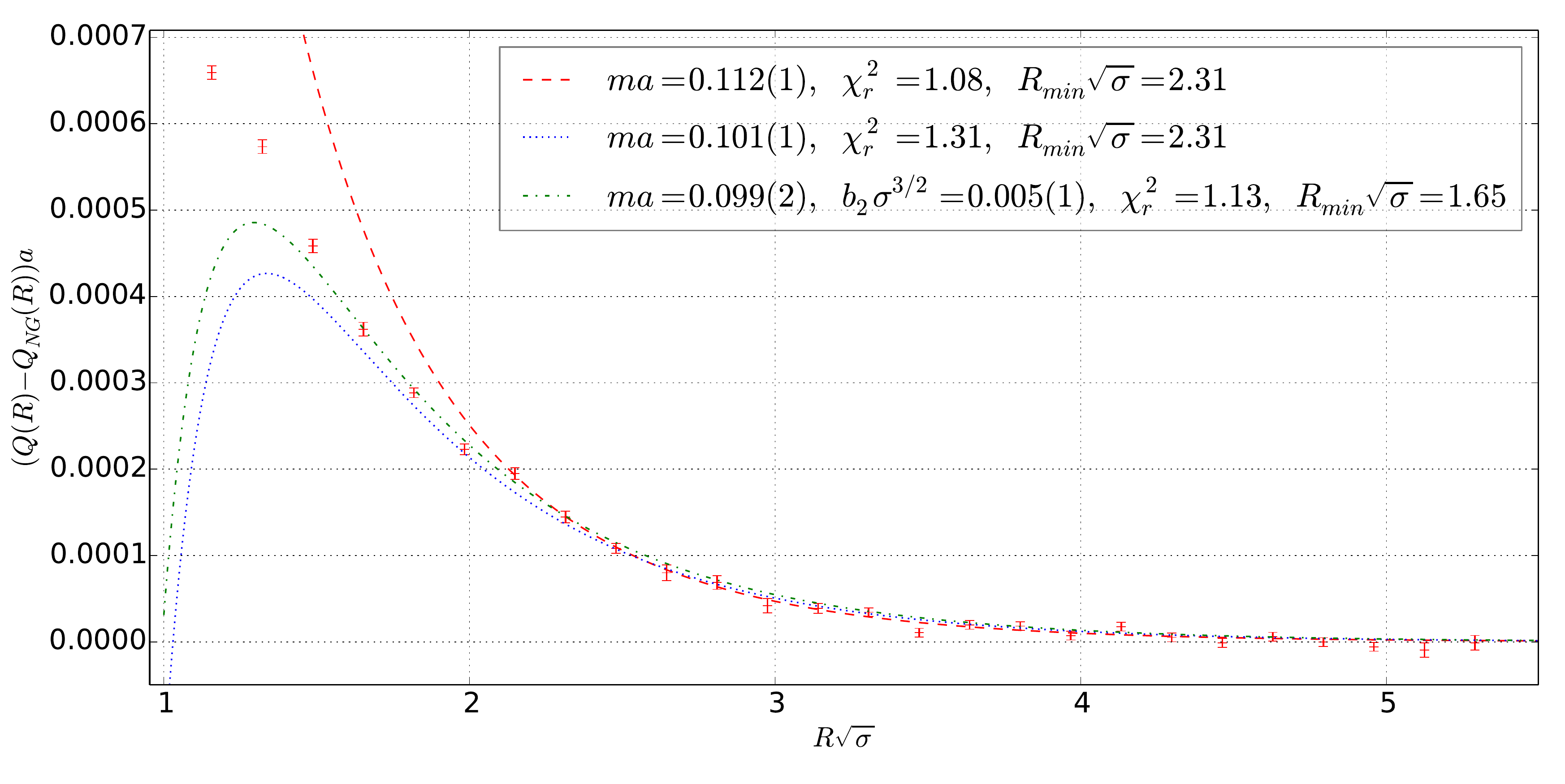}}
\caption{Fits of the $[Q-\QNG(R)]$ differences at $\beta=2.2$, with $\Vext$, $\Vext+V'_2$ and $\Vext+V'_2+$ boundary term.\label{fig:fig2}}
\end{figure}

We then tried to fit the $[Q(R)-\QNG(R)]$ differences with the rigid-string prediction $Q_r(R)$.\footnote{The sum over Bessel functions was truncated at $n=100$; the corresponding error was much smaller than the other uncertainties of our data.} In this case, fits with $m$ as the only free parameter successfully describe the data, even for small values of $\Rmin$, and the resulting $m$ values show the expected scaling behavior, $m \propto m_0$ (see tab.~\ref{tab:fit2}). For example, for $\beta=2.2$, a $\redchisq$ of order $1$ is obtained for 
$\Rmin\sqrt{\sigma}=2.15$: a clear improvement over the one-parameter fit to 
the pure Nambu-Goto model. Our results for $Q(R)$ at $\beta=2.2$ are shown in 
fig.~\ref{fig:fig1}, with the fitted curves.

\begin{table}
\begin{minipage}[b]{0.5\linewidth}
\centering
\begin{tabular}{|c|c|c|c|}
\hline
$\beta$  &$\sigma$  & $m_0$ & $L,N_t$\\
\hline
$1.7$ &$ 0.122764(2)$  & $0.88(1)$ & $64$\\
$1.9$ &$ 0.066824(6)$  & $0.56(1)$ & $64$\\
$2.0$ &$ 0.049364(2)$  & $0.44(1)$& $64$\\
$2.2$ &$ 0.027322(2)$  & $0.27(1)$ & $64$\\
$2.4$ &$ 0.015456(7)$  &$0.197(10)$& $128$\\
\hline
\end{tabular}
\caption{Information on the setup of our simulations.}
\label{tab:simul}
\end{minipage}
\begin{minipage}[b]{0.5\linewidth}
\centering
 \begin{tabular}{|c|c|c|c|}
 \hline
 $\beta$ & $m$ & $m_0$ & $m/m_0$\\
 \hline
 $1.7$ & $0.28(9)$ & $0.88(1)$ & $0.32(10)$\\
 $1.9$ & $0.25(4)$ & $0.56(1)$ & $0.45(7)$\\
 $2.0$ & $0.17(2)$ & $0.44(1)$& $0.39(4)$\\
 $2.2$ & $0.11(1)$ & $0.27(1)$ & $0.41(4)$\\
 $2.4$ & $0.06(2)$ &$0.20(1)$& $0.30(10)$\\
 \hline
 \end{tabular}
 \caption{Best-fit results for $m$ obtained using a three-parameter fit to our data, as explained in the text.}
 \label{tab:fit2}
\end{minipage}
\end{table}

We also tested if the NLO correction eq.~(\ref{eq:qrprime}) from 
the rigidity term could be seen in our lattice data, by fitting the $[Q(R)-\QNG(R)]$ differences, with $m$ as the only fitted parameter. This leads to better fits, especially for  $1<\Rmin\sqrt{\sigma}<2$, but the $\redchisq$ values are still larger 
than one. Moreover, the best-fit values for $m$ exhibit changes larger than their uncertainties. Therefore, this term cannot be neglected, within the precision of our lattice data. Note that this correction is of the same order as the boundary correction previously discussed. Hence it is possible that both boundary and rigid-string
corrections are present. To test this hypothesis, we set $m$ and $\sigma$ to the previously obtained best-fit values, and carried out a one-parameter fit of $[Q(R)-\QNG-Q'_r(R)]$ with the boundary correction $V_b(R)$, using $b_2$ as the free parameter. This leads to much better fits, including at small $\Rmin\sqrt{\sigma}$: $\redchisq$ values around one are obtained 
already for $\Rmin\sqrt{\sigma}\sim 1.65$, with a small but non-negligible value of $b_2\sigma^{3/2}=0.005(1)$. This shows that also this term is non-vanishing, and that it should be included in the analysis. However, disentangling the contributions from the boundary and the rigidity terms would require more precise data.

The $\Rmin$ dependence of the fits reveals that the rigid-string correction could affect the value of $\sigma$. Carrying out two-parameter fits of $Q(R)$ to the function $\QNG+Q_r(R)$ (or $\QNG+Q'_r(R)$) using $\sigma$ and $m$ as free parameters, we find that the best-fit value of $\sigma$ has a non-negligible impact on $m$ and on its uncertainty, and that including the NLO rigid-string correction changes the result for $m$.

It is clear that both the best-fit values of $\sigma$ and $b_2$ affect our estimate of $m$. In order to account for this interplay, we chose the result of a three-parameter fit to the data (using $\sigma$, $m$ and $b_2$ as free parameters) as our estimate for $m$. As $\sigma$ and $b_2$ are not fixed, this leads to a slightly larger error on $m$. We report these estimates for $m$ in tab.~\ref{tab:fit2}. The last column shows the $m/m_0$ ratio, which reveals good scaling properties. Our final estimate for the rigid-string parameter, including both statistical and systematic uncertainties into the error budget, is
\eq
\label{rigidstringparameterresult}
\frac{m}{m_0}=0.35(10).
\en

\section{Concluding remarks}
\label{sect:conclusions}

Our results show that in the 3D $\U(1)$ model, large deviations from the Nambu-Goto model appear for $\beta\geq2$. These deviations grow as $\beta$ is increased towards the continuum limit, and are accounted for by an extrinsic-curvature term in the effective string action. The rigidity parameter $m$ scales with the mass $m_0$ of the lightest glueball, and becomes dominant in the continuum limit, making this string very different from the Nambu-Goto one.

A necessary future step in our analysis is to disentangle the NLO contribution
of the rigidity term from the one due to the boundary term, and to study the fate of the rigid string at finite temperature. Another interesting extension of this work would be to study the behavior of the string width at large $\beta$, which should differ from the Nambu-Goto prediction.

{\bf Acknowledgments} This work is supported by the Spanish MINECO (grant FPA2012-31686 and ``Centro de Excelencia Severo Ochoa'' programme grant SEV-2012-0249).

\bibliography{paper}

\end{document}